\documentclass[te,nameyear,draft]{econsocart}

\makeatletter
\let\numberlines@hook\relax
\let\set@numberlines@box\relax
\let\put@numberlines@box\relax
\makeatother

\RequirePackage[colorlinks,citecolor=blue,linkcolor=blue,urlcolor=blue,pagebackref]{hyperref}

\startlocaldefs

\theoremstyle{plain}

\newtheorem{proposition}{Proposition}

\theoremstyle{definition}
\newtheorem{definition}{Definition}

\usepackage[capitalise]{cleveref}
\crefname{figure}{Figure}{Figures}
\Crefname{figure}{Figure}{Figures}
\crefname{assumption}{Assumption}{Assumptions}
\crefname{footnote}{footnote}{footnotes}

\usepackage{tikz}
\usepackage{tkz-euclide}
\usetikzlibrary{intersections}
\usepackage{pgfplots}
    \pgfplotsset{compat=1.15}
\usepackage{float}

\usepackage{mathtools}
\DeclarePairedDelimiter\ceil{\lceil}{\rceil}

\endlocaldefs

\makeatletter
\def\copyright@text{}
\def\@oddhead{\hfil\small\thepage}
\def\@evenhead{\small\thepage\hfil}
\makeatother

\begin{document}

\begin{frontmatter}

\title{Collective Experimentation with Correlated Payoffs}
\runtitle{Collective experimentation}

\begin{aug}
%
%
%
\author[add1]{\fnms{Kailin}~\snm{Chen}\ead[label=e1]{kailin.chen@aalto.fi}}
\address[add1]{%
\orgdiv{Department of Economics},
\orgname{Aalto University}}

\end{aug}

\begin{funding}
The author is indebted to Daniel Hauser, Johannes Hörner, Stephan Lauermann, Pauli Murto, Sven Rady, and Juuso Välimäki for invaluable feedback and support. For helpful discussions, the author thanks Simon Block, Carlo Cusumano, Francesc Dilme, Mehmet Ekmekci, Zhengqing Gui, Marina Halac, Mathijs Janssen, Zizhen Ma, Benny Moldovanu, Axel Niemeyer, Jorge Ramos-Mercado, Julia Salmi, Anna Sanktjohanser, Bruno Strulovici, Caroline Thomas, audiences at the European Winter Meeting of the Econometric Society, and anonymous referees at the ACM Conference on Economics and Computation (EC).
\end{funding}
\begin{abstract}
This paper studies an exponential bandit model in which a group of agents collectively decide whether to undertake a risky action $R$. This action is implemented if the fraction of agents voting for it exceeds a predetermined threshold $k$. Building on \citet{strulovici2010learning}, which assumes the agents' payoffs are independent, we explore the case in which the agents' payoffs are correlated. During experimentation, each agent learns individually whether she benefits from $R$; in this way, she also gains information about its overall desirability. Furthermore, each agent is able to learn indirectly from the others, because in making her decisions, she conditions on being pivotal (i.e., she assumes her vote will determine the collective outcome). We show that, when the number of agents is large, increasing the threshold $k$ for implementing $R$ leads to increased experimentation. However, information regarding the overall desirability of $R$ is effectively aggregated only if $k$ is sufficiently low.
\end{abstract}

\begin{keyword}
\kwd{Strategic experimentation}
\kwd{information aggregation}
\end{keyword}

\begin{keyword}[class=JEL] 
\kwd{C73}
\kwd{D70}
\kwd{D83}
\end{keyword}

\end{frontmatter}

\section{Introduction}
\label{sec:1}

Strategic experimentation and continuous learning play a critical role in many types of real-world decision-making. For example, a corporation may introduce a new product to the market without knowing whether it will be profitable. Over time, the directors of the corporation's divisions can gauge the success of the product and then decide jointly whether to continue producing it. Similarly, citizens of a country may need to vote on a political reform without being able to fully anticipate its effects. After the initial implementation of the reform, the citizens will gradually learn whether it benefits them, i.e., whether they are ``winners'' or ``losers'' under it. They can then vote either to continue the reform or to revert to the status quo.\footnote{\citet{fernandez1991resistance} and \citet{ali2025political} show that in static settings, asymmetric uncertainty regarding the distribution of gains and losses from the reform induces a status-quo bias.}

One canonical framework for studying strategic experimentation and continuous learning with multiple agents is the exponential bandit model analyzed by \citet{keller2005strategic}. In that model, each agent decides individually whether to undertake a certain risky action. However, in many real-world situations (such as the examples above), a group of agents must jointly decide on a collective action. The collective decision-making process creates incentives for experimentation that are qualitatively different from those of \citet{keller2005strategic}. In particular, even when the agents have ex ante common interests, the information they gain from experimentation may cause their preferences to diverge and alter the distribution of gains and losses among them.

The seminal paper of \citet{strulovici2010learning} embeds collective decision‑making into the framework of \citet{keller2005strategic} and investigates the inefficiencies generated by diverging preferences. \citeauthor{strulovici2010learning} shows that collective decision-making leads to substantial conservatism, so that equilibrium experimentation with the risky action falls below the socially efficient level. This occurs because each agent fears being trapped into an unfavorable course of action by the future choices of other agents after they receive new information from experimentation---a phenomenon \citet{strulovici2010learning} explains in terms of ``winner frustration'' and the ``loser trap''.



A critical assumption in the main model of \citet{strulovici2010learning} is that each agent learns whether the risky action benefits her only from her own payoff stream, and the success or failure of other agents yields no informational value for her; that is, the agents' payoffs are independent. In this paper, we revisit \citeauthor{strulovici2010learning}'s model under the assumption that the agents' payoffs are correlated. This assumption holds in many important applications. For example, citizens voting on a reform typically have correlated payoffs that depend on the overall desirability of the reform. Thus, if a citizen knows whether others have benefited from the reform, she has some information about whether she will benefit from it as well.

Another important aspect of the decision-making process is the extent to which agents can observe each other's payoffs. \citet{keller2005strategic} assume that payoffs are publicly observed. However, in many applications, the agents may not be able to directly access or accurately assess other agents' payoffs. \citet{strulovici2010learning} analyzes the case with publicly observed payoffs as a benchmark. He then shows that when payoffs are privately observed, the agents can implement the same collective decision policy as they would in the publicly observed case, provided that a reversible and costless action is available. The agents switch to this action at specific times according to their payoffs. As a result, each agent can learn about the others' payoffs from the past collective outcomes. 

In this paper, we assume privately observed payoffs and identify a novel channel for learning: Each agent makes her decisions while conditioning on being pivotal (i.e., conditioning on the event that her vote will determine the collective outcome), and this allows her to deduce some information about the other agents' payoffs. Exactly what she is able to learn depends on the collective decision rule in place. In our model, we consider a qualified majority rule under which the risky action is implemented if the fraction of agents supporting it exceeds a predetermined threshold. We show that a higher threshold creates stronger incentives for experimentation, because whenever such a threshold is met, each agent infers that there are more other agents already benefiting from the risky action, which increases her optimism about benefiting from it herself. Surprisingly, however, when the threshold becomes too large, it leads to over-experimentation that generates inefficiency.

Specifically, we analyze a two-armed exponential bandit model involving $N$ agents. At each instant, the agents jointly decide between a safe action $S$ and a risky action $R$ under a predetermined collective decision rule: Each agent casts a vote for either $R$ or $S$. If the share of votes for $R$ exceeds a given threshold $k\in(0,1]$, then $R$ is implemented; otherwise, $S$ is implemented. The safe action $S$ yields a constant, homogeneous payoff to each agent, while $R$ yields intermittent payoffs depending on the agents' types, which are unknown initially. If an agent's type is bad, then $R$ pays her nothing. If her type is good, then $R$ pays her positive lump-sum payoffs at random times corresponding to the arrival times of a Poisson process. Thus, if an agent receives a lump sum, then she learns that her type is good; in this case, we call her a \emph{sure winner}. An agent who has not yet received a lump sum is still uncertain about her type and is called an \emph{unsure voter}. Unsure voters become increasingly pessimistic about their type with time. 

As stated earlier, we assume that payoffs are privately observed; that is, each agent observes only her own payoff stream. We also assume the agents' types are correlated: Before the game starts, nature chooses at random a state $\omega\in\{H,L\}$, which is unknown to the agents. Nature then chooses each agent's type independently. An agent's type is more likely to be good in state $H$ than in state $L$. Hence, when there are more sure winners, the unsure voters have stronger incentives to experiment (i.e., to choose $R$): Each unsure voter believes the state is more likely to be $H$, and so her type is more likely to be good.

In our main analysis, we assume the safe action $S$ is irreversible; that is, once implemented, it remains fixed. For this setting, we show that there exists a unique symmetric pure-strategy equilibrium in undominated strategies. In the equilibrium, sure winners always vote for $R$, while all unsure voters adopt the same cut-off strategy: They vote for $R$ at all instants up to a cut-off time $\hat{t}$ and for $S$ thereafter. At time $\hat{t}$, they are indifferent between $R$ and $S$. 

Each unsure voter at time $\hat{t}$ makes her decision conditional on being pivotal (i.e., on being capable of changing the collective outcome). More precisely, she updates her belief about the realized state conditional on the event that there are exactly $kN-1$ sure winners, and then forms her belief regarding her own type. Therefore, different collective decision rules yield different equilibrium cut-off times $\hat{t}$. Strategic voting thus affects the agents' incentives for experimentation.

Our main results identify the limiting properties of the equilibrium cut-off time as the number of agents grows large. We find that the limit cut-off time is increasing in $k$; that is, when the threshold for implementing $R$ is higher, the agents are willing to experiment for longer. However, the limit cut-off time is bounded above by the stopping time chosen by a myopic decision-maker who is certain that the realized state is $H$. The intuition behind these results is as follows. When the number of agents is large, each unsure voter conditions on the event that she is pivotal at the equilibrium cut-off time, but never thereafter, because (in the limit) individual control over future decisions becomes infinitely diluted. Consequently, unsure voters behave myopically. Moreover, when $k$ is higher, unsure voters become more optimistic that the realized state is $H$ and their type is good, because they infer from being pivotal that there are more sure winners. Hence a higher $k$ leads to increased experimentation. However, when unsure voters are already certain that the realized state is $H$, increasing $k$ cannot further increase their optimism about their type. This implies the stated upper bound on the limit cut-off time.

We also examine whether, in the limit, the dispersed information gained by the agents through experimentation is effectively aggregated. We assume that $R$ yields a higher expected flow payoff than $S$ in state $H$, and $S$ yields a higher expected flow payoff than $R$ in state $L$, and say that information is aggregated if, after the limit cut-off time, $R$ is always implemented in state $H$ and $S$ in state $L$.

We find that information is aggregated whenever $k$ falls below a certain critical value. Therefore, to enable information aggregation, the collective decision rule should favor experimentation. This result emphasizes a key trade-off inherent in collective experimentation: While it generates new information about the true state, it also creates heterogeneity in the agents’ beliefs about their own types, which impairs collective decision-making. More precisely, as previously discussed, experimentation increases (i.e., agents wait longer before switching to $S$) as $k$ increases. When $k$ is small, the agents’ beliefs are relatively homogeneous at the equilibrium cut-off time: Sure winners prefer $R$ in both states, while unsure voters prefer $R$ in state $H$ and $S$ in state $L$. Now, when the number of agents is sufficiently large, the vote share of $R$ at the equilibrium cut-off time must be greater than $k$ in state $H$ and less than $k$ in state $L$. If it were not, then each unsure voter---conditional on being pivotal---would infer that the realized state is that in which the vote share of $R$ is closer to $k$, which precludes indifference between $R$ and $S$ at the equilibrium cut-off time. The argument concerning information aggregation is similar to that used for the static voting model of \citet{feddersen1997voting} and \citet{duggan2001bayesian}, in which strategic voters make inferences conditional on being pivotal and attempt to match policies to states. 
 
However, when $k$ is large, the agents over-experiment, and their beliefs become highly heterogeneous at the equilibrium cut-off time: Sure winners prefer $R$ in both states, while unsure voters prefer $S$ in both states. In particular, the unsure voters whose type is good---and who would therefore vote for $R$ if they knew their type---develop pessimistic beliefs and so vote for $S$ instead. In this way, the heterogeneity in the agents' beliefs induces a bias towards $S$, and so information is not aggregated.  This phenomenon relates our work to that of \citet{fernandez1991resistance} and \citet{ali2025political}. These papers show that asymmetric uncertainty impairs collective decision-making within a static model in which both the collective decision rule and the asymmetric uncertainty are exogenously imposed. In our paper, by contrast, asymmetric uncertainty arises endogenously from the collective decision rule.

To conclude our analysis, we consider a variation of our main model in which the safe action $S$ is reversible, so that the agents can learn from the previous collective outcomes. The equilibrium identified in our main model remains valid in this setting. Furthermore, we show that when $k$ is above the critical value established in the irreversible case, there is no equilibrium in undominated strategies that aggregates information. This finding reinforces our argument that the collective decision rule should favor experimentation.

The paper proceeds as follows. In \cref{sec:2} we describe the model and characterize the equilibrium. In \cref{sec:3} we analyze the limit as the number of agents grows large and present our result on information aggregation. In \cref{sec:4} we consider the modified setting in which the safe action is reversible. In \cref{sec:5} we survey the related literature. \cref{sec:6} concludes.


\section{Model and Equilibrium}
\label{sec:2}

\subsection{Model Setting}
\label{sec:2.1}
We study an exponential bandit model in continuous time with $t\in[0,\infty)$. Payoffs are discounted at rate $r$. There is a set of agents, denoted by $\{1,\ldots,N\}$ with $N\geq 1$. The agents vote continuously over time between two actions, a safe action $S$ and a risky action $R$. There is a fixed \emph{threshold} $k\in(0,1]$, such that at each time $t\in[0,\infty)$, the risky action $R$ is implemented if the number of votes for $R$ is at least $kN$. (For convenience, we assume $kN$ is an integer.\footnote{This assumption is unnecessary but simplifies the exposition. Equivalently, we could drop the assumption and suppose $R$ is implemented if and only if at least $\ceil*{kN}$ agents vote for $R$.}) If at any time $t$ there are fewer than $kN$ votes for $R$, then the game ends, and the safe action $S$ is implemented for all subsequent periods.\footnote{In other words, the safe action $S$ is irreversible; the risky action, once rejected, cannot be restarted. This setting corresponds to situations in which restarting costs are prohibitively high. In Section \ref{sec:4} we consider a variant of the model in which $S$ is reversible.} The number of votes that have been cast for each alternative is always public.

Before the game starts, nature chooses a state randomly from $\{H,L\}$. The agents are uncertain about the state; they hold a common prior belief $q_0$ that the state is $H$. After choosing the state, nature chooses the type of each agent independently. Each agent's type is either good or bad. An agent's type is good with probability $\rho_H$ if the state is $H$, and with probability $\rho_L$ if the state is $L$. We assume $\rho_H>\rho_L> 0$; that is, each agent's type is more likely to be good in state $H$ than in state $L$.\footnote{Many of our results, including Propositions \cref{prop:1}, \cref{prop:3}, and \cref{prop:4}, can be extended to the case $\rho_L=0$.} All types are initially unobservable to all agents.\footnote{\citet{strulovici2010learning} analyzes the case of two agents with correlated payoffs, with unanimity required for the risky action to be implemented. We explore the effect of correlated payoffs in a more general setting, with an arbitrary number of agents and a more flexible collective decision rule.}

If the safe action $S$ is implemented, it yields a flow $s$ per unit of time to all agents. If the risky action $R$ is implemented, each agent's payoff depends on her type. If her type is bad, then $R$ always pays her $0$. If her type is good, then $R$ pays her positive lump sums, each of magnitude $z$, at random times corresponding to the arrival times of a Poisson process with constant intensity $\lambda>0$. The arrival times of the lump sums are independent among the agents. We denote by $g=\lambda z$ the expected payoff per unit of time from $R$ for an agent with the good type. Payoffs are privately observed; that is, each agent observes only her own payoff stream.

We assume that
\begin{equation}
\label{eqn:1}
    \rho_H g>s>\rho_L g> 0.
\end{equation}
That is, an agent's expected flow payoff from $R$ (based on her prior belief about her type) is higher than that from $S$ in state $H$, but lower in state $L$. For ease of exposition, we further assume that
\begin{equation}
\label{eqn:2}
    q_0\rho_H g+(1-q_0)\rho_L g>s,
\end{equation}
i.e., $R$ yields a higher expected flow payoff than $S$ under an agent's prior beliefs about the state and her type.

At each time $t\in[0,\infty)$, the agents can be divided into two groups: the \emph{sure winners}, who have received lump-sum payments before time $t$ and are therefore certain their type is good, and the \emph{unsure voters}, who have not yet received lump sums. Each unsure voter assigns the same probability to her type being good.

\subsection{Equilibria}
\label{sec:2.2}

We consider perfect Bayesian equilibria in which (\romannumeral 1) sure winners always vote for $R$, and (\romannumeral 2) all unsure voters use the same pure strategy, unanimously voting either for $R$ or for $S$. Since the safe action $S$ is irreversible, an equilibrium is characterized by a cut-off time $\hat{t}_{k,N}\geq 0$ such that unsure voters vote for $R$ when $t<\hat{t}_{k,N}$ and for $S$ when $t\geq\hat{t}_{k,N}$.\footnote{In \cref{sec:2.3} we characterize the on-path beliefs of the unsure voters. As for their off-path beliefs, when an unsure voter observes votes for $S$ before $\hat{t}_{k,N}$, she infers that such votes originate from other unsure voters.} An equilibrium thus proceeds as follows: at each time $t<\hat{t}_{k,N}$, all the agents vote for $R$, leading to its implementation. At $\hat{t}_{k,N}$, if there are at least $kN$ sure winners, then $R$ is implemented for all subsequent periods. Otherwise, $S$ is implemented for all subsequent periods. 

Consider an equilibrium with $\hat{t}_{k,N}>0$. Suppose that at time $\hat{t}_{k,N}$ there is an unsure voter. From this voter's perspective, if there are more than $kN-1$ sure winners at time $\hat{t}_{k,N}$, then $R$ is implemented for all subsequent periods regardless of her own vote. If there are fewer than $kN-1$ sure winners, then $S$ is implemented for all subsequent periods. Hence the unsure voter's vote affects the collective outcome only if there are exactly $kN-1$ sure winners. It follows that at time $\hat{t}_{k,N}$, each unsure voter makes her decision conditional on the event that there are exactly $kN-1$ sure winners, i.e., that she is \emph{pivotal} for the collective outcome. (In this event we also say she has \emph{full control of experimentation}.) 

In particular, even though the agents do not observe each other's payoffs, they can still infer the number of sure winners and gain information about the aggregate state conditional on being pivotal.

We analyze equilibria in undominated strategies, with the following requirements: (\romannumeral 1) $\hat{t}_{k,N}>0$, and (\romannumeral 2) each unsure voter is indifferent between $R$ and $S$ at $\hat{t}_{k,N}$ conditional on being pivotal. The assumption \eqref{eqn:2} from the previous section rules out the equilibrium with $\hat{t}_{k,N}=0$, i.e., the equilibrium in which $S$ is always implemented. As for the requirement (ii), consider an equilibrium with a cut-off time $\hat{t}$ at which each unsure voter strictly prefers $S$ to $R$ (conditional on being pivotal). Then we can find $\hat{t}'<\hat{t}$ such that every unsure voter would prefer the strategy profile in which all of the unsure voters switch to $S$ at $\hat{t}'$, instead of waiting until $\hat{t}$, regardless of the number of sure winners at $\hat{t}'$. Furthermore, if at $\hat{t}'$ there are fewer than $kN$ sure winners (i.e., if the unsure voters can change the collective outcome), then the unsure voters \emph{strictly} prefer to switch to $S$ at $\hat{t}'$. Thus, from the perspective of a given unsure voter, it would be ideal if all of the unsure voters voted for $S$ at $\hat{t}'$ rather than at $\hat{t}$. However, since in this equilibrium all of the other unsure voters vote for $R$ at $\hat{t}'$, her vote is non-pivotal and so she also votes for $R$ at $\hat{t}'$. We omit this equilibrium from consideration as it results from a trivial coordination issue in the collective decision problem.\footnote{We can rule out this equilibrium through a refinement based on the elimination of conditionally dominated strategies, similarly to \citet{strulovici2010learning}, which requires each unsure voter to vote as if she were pivotal at each instant. We can also rule out the equilibrium by a trembling-hand argument: Let each unsure voter employ a mixed strategy, following the cut-off strategy $\hat{t}$ with probability $1-\epsilon$ and the cut-off strategy $\hat{t}-\delta$ with probability $\epsilon$. Then there exists some $\delta>0$ such that, regardless of the value of $\epsilon$, each unsure voter strictly prefers to vote for $S$ at time $\hat{t}-\delta$.} 

Henceforth, an equilibrium in undominated strategies is referred to simply as an equilibrium.

\subsection{Beliefs}
 \label{sec:2.3}

We now examine the agents' beliefs at a given time $t$. Suppose $R$ has been implemented at all times before $t$. Then an unsure voter's belief that her type is good, conditional on the true state being $H$ or $L$ respectively, is as follows:
\begin{eqnarray*}
    Pr(good|H,t)&=&\frac{\rho_H e^{-\lambda t}}{\rho_H e^{-\lambda t}+1-\rho_H},\\
    Pr(good|L,t)&=&\frac{\rho_L e^{-\lambda t}}{\rho_L e^{-\lambda t}+1-\rho_L}.
\end{eqnarray*}
Note that in each state, the probability that an agent of the good type has not yet received a lump sum is $e^{-\lambda t}$.

Next, we calculate the probability that the state is $H$ (or that it is $L$), conditional on the event that there are $K$ sure winners at $t$. In state $\omega$, the probability that an agent has received a lump sum before time $t$ is $\rho_{\omega}(1-e^{-\lambda t})$. Hence
\begin{equation}
\label{eqn:3}
    \frac{Pr(H|K,t)}{Pr(L|K,t)}=\underbrace{\frac{q_0}{1-q_0}}_{\text{prior}}
    \underbrace{\left[\frac{\rho_H(1-e^{-\lambda t})}{\rho_L(1-e^{-\lambda t})}\right]^{K}}_{K\text{ sure winners}}
    \underbrace{\left[\frac{1-\rho_{H}(1-e^{-\lambda t})}{1-\rho_{L}(1-e^{-\lambda t})}\right]^{N-K}}_{N-K\text{ unsure voters}}.
\end{equation}

Finally, we calculate an unsure voter's belief that her type is good conditional on the event that there are $K$ sure winners at $t$:
\begin{equation}
\label{eqn:4}
    Pr(good|K,t)=\underbrace{Pr(H|K,t)Pr(good|H,t)}_{\text{state $H$ and good type }}+\underbrace{Pr(L|K,t)Pr(good|L,t)}_{\text{state $L$ and good type }}.
\end{equation}

Both $Pr(good|H,t)$ and $Pr(good|L,t)$ are strictly decreasing in $t$. Both $Pr(H|K,t)$ and $Pr(good|K,t)$ are strictly increasing in $K$ and strictly decreasing in $t$.

\subsection{Equilibrium Characterization}
\label{sec:2.4}

We now calculate the equilibrium cut-off time $\hat{t}_{k,N}$. Consider an agent $i$ who is an unsure voter at time $\hat{t}_{k,N}$. As observed above, her decisions are conditional on the event that she is pivotal (i.e., that there are exactly $kN-1$ sure winners) at time $\hat{t}_{k,N}$. She thus believes that her type is good with probability $Pr(good|kN-1,\hat{t}_{k,N})$. Since she is indifferent between voting for $R$ and voting for $S$, the equilibrium cut-off time $\hat{t}_{k,N}$ must satisfy
 \begin{equation}
 \label{eqn:5}
    \begin{aligned}
    s={}&\underbrace{Pr(good|kN-1,\hat{t}_{k,N})g}_{\text{flow payoff from $R$}}+\underbrace{Pr(good|kN-1,\hat{t}_{k,N})\lambda\left(\frac{g}{r}-\frac{s}{r}\right)}_{\text{jump when agent $i$ receives a lump sum}}\\
    &+\underbrace{(N-kN)Pr(good|kN-1,\hat{t}_{k,N})\lambda\left[\frac{Pr(good|kN,\hat{t}_{k,N})g}{r}-\frac{s}{r}\right]}_{\text{jump when another unsure voter receives a lump sum}}.
    \end{aligned}
\end{equation}
In words, the interpretation of this equation is as follows. Conditional on being pivotal, agent $i$ has full control of experimentation: her voting for either $R$ or $S$ implements that choice for the next instant. The left-hand side of equation \eqref{eqn:5} is the flow payoff generated by $S$. The right-hand side is the sum of the expected flow payoff from $R$ and two possible payoff jumps: one when agent $i$ receives a lump sum, and another when other unsure voters receive lump sums. If agent $i$ receives a lump sum in the next instant, which happens with probability $Pr(good|kN-1,\hat{t}_{k,N})\lambda dt$, she becomes a sure winner and $R$ is implemented for all subsequent periods. Her expected payoff thus jumps from $\frac{s}{r}$ to $\frac{g}{r}$. If another unsure voter receives a lump sum in the next instant, which happens with probability $(N-kN)Pr(good|kN-1,\hat{t}_{k,N})\lambda dt$, then $R$ is implemented for all subsequent periods while agent $i$ remains an unsure voter with belief $Pr(good|kN,\hat{t}_{k,N})$. Her expected payoff jumps from $\frac{s}{r}$ to $\frac{Pr(good|kN,\hat{t}_{k,N})g}{r}$.

\begin{proposition}
\label{prop:1}
For each $k\in(0,1]$ and $N\geq 1$, a unique equilibrium exists.
\end{proposition}

Note that when the right-hand side of \eqref{eqn:5} is positive, it is strictly decreasing in $\hat{t}_{k,N}$, because both $Pr(good|kN-1,\hat{t}_{k,N})$ and $Pr(good|kN,\hat{t}_{k,N})$ are strictly decreasing in $\hat{t}_{k,N}$. Furthermore, it converges to $0$ as $\hat{t}_{k,N}\to\infty$, since
\begin{equation*}
    \lim_{t\to\infty}Pr(good|kN-1,t)=0\,\;\forall k\in(0,1), N\geq 1.
\end{equation*}
Finally, by \eqref{eqn:2},\footnote{If \eqref{eqn:2} is violated, there exist $k^*\in(0,1)$ and $N^*(k)\geq 0$ such that for each $k\in(0,k^*)$, when $N>N^*(k)$, there is a unique equilibrium satisfying $\hat{t}_{k,N}>0$.} the right-hand side of \eqref{eqn:5} converges to a value greater than $s$ as $\hat{t}_{k,N}\to0$. Therefore, by the intermediate value theorem, there exists a unique solution to \eqref{eqn:5}.


\section{Large Number of Agents}
\label{sec:3}

In this section, we examine the limiting properties of the sequence of equilibria as the number of agents $N$ grows large.

\subsection{Limit Cut-Off}
\label{sec:3.1}

Here we characterize the limit of the equilibrium cut-offs $\{\hat{t}_{k,N}\}$ as $N\to\infty$. Let $\overline{t}$ be the unique solution to 
\begin{equation*}
    Pr(good|H,\overline{t})=\frac{s}{g}.
\end{equation*}
We interpret this time $\overline{t}$ as follows. Consider a version of our game involving only a single agent, who chooses between $R$ and $S$ at each instant. Assume that (\romannumeral 1) this agent is certain that the state is $H$ but uncertain about her type, and (\romannumeral 2) she is myopic: she cares only about the flow payoff. Then $\overline{t}$ is the time at which, if she has not yet received a lump sum, she starts choosing $S$ rather than $R$.

\begin{proposition}
\label{prop:2}
For each $k\in(0,1]$, there exists $\hat{t}_{k}\in(0,\bar{t}\,]$ such that
\begin{equation*}
    \lim_{N\to\infty}\hat{t}_{k,N}=\hat{t}_{k}.
\end{equation*}
There exists $\overline{k}\in(0,1)$ such that when $k\leq\overline{k}$, the limit cut-off $\hat{t}_{k}$ is strictly increasing in $k$; furthermore, $\lim_{k\to0}\hat{t}_{k}=0$ and $\hat{t}_{\overline{k}}=\overline{t}$. When $k>\overline{k}$, the limit cut-off $\hat{t}_{k}$ is equal to the myopic cut-off $\overline{t}$.\footnote{A similar statement holds when $\rho_L=0$: In that case, $\hat{t}_k=\bar{t}$ for every $k\in(0,1]$. This is because each unsure voter realizes that, conditional on her being pivotal, there must be at least one sure winner, and hence the state must be $H$.}
\end{proposition}

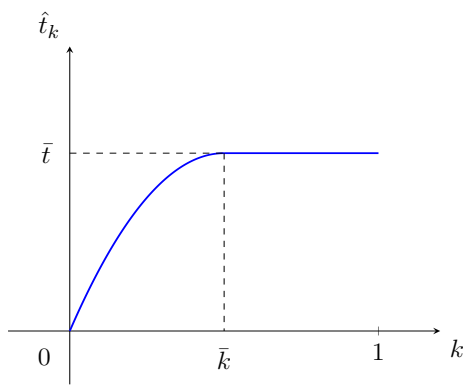
\begin{figure}[H]
\centering
\resizebox{2.5in}{2.0in}{
\begin{tikzpicture}
\begin{axis}[
        xmin=-0.2,
        xmax=1.2,
        ymin=-0.3,
        ymax=1.6,
        axis lines=center,
        xtick={0,1},
        ytick={\empty},
        xlabel={$k$},
        ylabel={$\hat{t}_k$},
        xlabel style={below right},
        ylabel style={above left},
        smooth,
        domain=0:1,
        samples=100,
        no markers,
]
\node at (0,0) [label=below left:$0$] {};
\addplot [domain=0:0.5, color=blue, line width=0.3mm, name path global=function1]{1-4*(0.5-x)^2};
\addplot [domain=0.5:1, color=blue, line width=0.3mm, name path global=function2]{1};
\addplot[domain=0:0.5, color=black, line width=0.1mm, dashed, name path global=function3]{1}; 
\node at (0,1) [label=left:$\bar{t}$] {};
\draw [dashed]
        (0.5,1)          node                   {}
            -- (0.5,1 |-0,0)   node [label=below:$\bar{k}$] {};
\end{axis}
\end{tikzpicture}
}
\caption{The limit cut-off $\hat{t}_k$. When $k\leq\bar{k}$, $\hat{t}_k$ is strictly increasing in $k$: The agents are more willing to experiment when the threshold for $R$ to be implemented is higher. When $k>\bar{k}$, $\hat{t}_k=\bar{t}$: Each unsure voter behaves like a single myopic decision-maker who believes the realized state is $H$.}
    \label{fig:1}
\end{figure}

\cref{fig:1} illustrates \cref{prop:2}. 

For the intuition behind \cref{prop:2}, consider the perspective of a fixed agent $i$. Suppose that at a certain time $\hat{t}$, all of the agents other than $i$ who are unsure voters switch from $R$ to $S$. Suppose also that agent $i$ is an unsure voter and pivotal at $\hat{t}$. If agent $i$ votes for $R$ throughout the interval $[\hat{t},\hat{t}+dt)$, and an unsure voter receives a lump sum sometime during that interval, then $R$ will be implemented for all subsequent periods. However, when the number of agents is large, it is highly likely that the unsure voter receiving the lump sum will be an agent other than $i$, and so agent $i$ will cease to be pivotal. Thus, when the number of agents is large, agent $i$ condtions on the event that she is pivotal at time $\hat{t}$ and never thereafter, i.e., that she has full control of experimentation at time $\hat{t}$ and will lose control permanently in the next instant. Therefore, at time $\hat{t}$, agent $i$ behaves like a single myopic decision-maker who can choose to implement either $R$ or $S$ for all subsequent periods.

Since unsure voters condition on being pivotal, i.e., on the event that there are $kN-1$ sure winners, they are more optimistic that the realized state is $H$ (and thus that their type is good) as $k$ increases. Hence a higher threshold $k$ leads to a higher limit cut-off time $\hat{t}_k$. However, this effect is limited. Even if an unsure voter is certain that the realized state is $H$, after time $\bar{t}$ she votes for $S$ since she behaves myopically. Therefore $\hat{t}_k \leq \bar{t}$ for each $k\in(0,1]$.

\subsection{Information Aggregation}
\label{sec:3.2}

As long as the risky action $R$ is being implemented, the agents gradually learn their types and so gain dispersed information about the aggregate state. We now examine whether this dispersed information is effectively aggregated and utilized as the number of agents grows large. 

From \eqref{eqn:1}, $R$ yields a higher expected flow payoff than $S$ in state $H$, while the reverse is true in state $L$. Consider a utilitarian social planner who knows both the state and each agent's type. Under the planner's optimal decision, as $N\to\infty$, the probability that $R$ (resp.~$S$) is implemented in state $H$ (resp.~state $L$) converges to 1. We say a sequence of strategy profiles aggregates information if it leads to the same outcomes after experimentation.
 \begin{definition}
 \label{def:1}
       A sequence of strategy profiles for the game with $N$ agents, for each value of $N$ \emph{aggregates information} if for each $\epsilon>0$ there exist $N_\epsilon$ and $\hat{t}_{\epsilon}$ such that for each $N>N_\epsilon$, the following hold: 
        \begin{enumerate}
            \itemsep0.5em
            \item The event that, in state $H$, $R$ is implemented at each time $t\in[\hat{t}_{\epsilon},\infty)$ occurs with probability greater than $1-\epsilon$.
            \item The event that, in state $L$, $S$ is implemented at each time $t\in[\hat{t}_{\epsilon},\infty)$ occurs with probability greater than $1-\epsilon$. 
        \end{enumerate}
\end{definition}  

Let $H(\hat{t}_{k,N})$ and $L(\hat{t}_{k,N})$ denote the numbers of sure winners at the equilibrium cut-off time $\hat{t}_{k,N}$ in state $H$ and state $L$, respectively. Then the sequence of equilibria aggregates information if
        \begin{eqnarray*}
        \lim_{N\to\infty}Pr(H(\hat{t}_{k,N})>kN)=1,\\
        \lim_{N\to\infty}Pr(L(\hat{t}_{k,N})<kN)=1.
        \end{eqnarray*}

\begin{proposition}
\label{prop:3}
The sequence of equilibria aggregates information when $k\in(0,\frac{\rho_H g-s}{g-s})$.\footnote{If \eqref{eqn:2} is violated, a modified version of this statement holds: There is some $k^*\in(0,\frac{\rho_H g-s}{g-s})$ such that there exists a sequence of equilibria aggregating information if and only if $k<k^*$.} When $k\in[\frac{\rho_H g-s}{g-s},1]$, information is not aggregated, as
\begin{equation*}
    \lim_{N\to\infty}Pr(H(\hat{t}_{k,N})>kN)<1.
\end{equation*}
Furthermore, when $k\in(\frac{\rho_H g-s}{g-s},1]$, the probability of the event that $S$ is implemented at each time $t\in(\hat{t}_{k,N},\infty)$ converges to $1$ as $N\to\infty$.
\end{proposition}

\cref{prop:3} says that the sequence of equilibria aggregates information whenever $k$ is sufficiently low. For comparison, if the agents could observe their types before the start of the game, then information would be aggregated whenever $k\in(\rho_L,\rho_H)$. The critical threshold  $\frac{\rho_H g-s}{g-s}$ is strictly lower than $\rho_H$, and may even fall below $\rho_L$.

\cref{prop:3} emphasizes a key trade-off inherent in collective experimentation: While it generates new information about the true state, it also creates heterogeneity in the agents’ beliefs about their types, which impairs collective decision-making. More precisely, recall that by \cref{prop:2}, experimentation increases (i.e., agents wait longer before switching to $S$) as $k$ increases. When $k\leq\bar{k}$, the agents’ beliefs at $\hat{t}_k$ are relatively homogeneous: Sure winners prefer $R$ in both states, while unsure voters prefer $R$ in state $H$ and $S$ in state $L$. In this situation, \cref{prop:3} shows that strategic voting facilitates information aggregation. On the other hand, when $k>\bar{k}$, the agents over-experiment, and their beliefs at $\hat{t}_k$ become highly heterogeneous: sure winners prefer $R$ in both states, while unsure voters prefer $S$ in both states. In particular, the unsure voters whose type is good---and who would therefore vote for $R$ if they knew their type---develop pessimistic beliefs and so vote for $S$ instead. Thus, the heterogeneity in the agents' beliefs induces a bias towards $S$ and leads to the failure of information aggregation.

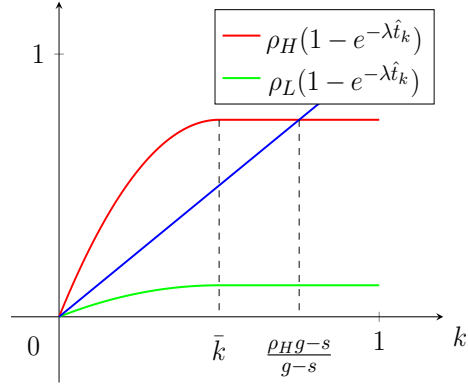
\begin{figure}[H]
\centering
\resizebox{2.5in}{2.0in}{
\begin{tikzpicture}
\begin{axis}[
        xmin=-0.15,
        xmax=1.2,
        ymin=-0.25,
        ymax=1.2,
        axis lines=center,
        xtick={0,1},
        ytick={0,1},
        xlabel={$k$},
        ylabel={},
        xlabel style={below right},
        ylabel style={above left},
        smooth,
        domain=0:1,
        samples=100,
        no markers,
]
\node at (0,0) [label=below left:$0$] {};
\addplot [domain=0:0.5, color=red, line width=0.3mm, name path global=function2]{(0.75)-3*(0.5-x)^2};
\addlegendentry{$\rho_{H}(1-e^{-\lambda \hat{t}_{k}})$}
\addplot [domain=0:0.5, color=green, line width=0.3mm, name path global=function2]{(0.12)-0.48*(0.5-x)^2};
\addlegendentry{$\rho_{L}(1-e^{-\lambda \hat{t}_{k}})$}
\addplot[domain=0.5:1, color=red, line width=0.3mm, name path global=function3]{0.75}; 
\addplot [domain=0.5:1, color=green, line width=0.3mm, name path global=function2]{0.12};
\addplot [domain=0:1, color=blue, line width=0.3mm, name path global=function1]{x};
\draw [dashed]
        (0.5,0.75)          node                   {}
            -- (0.5,0.75 |-0,0)   node [label=below:$\bar{k}$] {};
            
\draw [dashed]
        (0.75,0.75)          node                   {}
            -- (0.75,0.75 |-0,0)   node [label=below:$\frac{\rho_H g-s}{g-s}$] {};
\end{axis}
\end{tikzpicture}
}
\caption{The fraction of sure winners in each state. The red curve gives the fraction of sure winners at time $\hat{t}_k$ in state $H$, for all $k\in(0,1)$; the green curve gives the corresponding fraction for state $L$. By \cref{prop:2}, these quantities are strictly increasing in $k$ when $k<\bar{k}$, and are constant when $k\geq\bar{k}$. The blue line is the 45-degree line. The sequence of equilibria aggregates information for each $k$ at which the red curve lies above the blue line while the green curve lies below it.}
\label{fig:2}
\end{figure}

Let us sketch the proof of \cref{prop:3}. By the law of large numbers and \cref{prop:2}, the vote share of $R$ (i.e., the fraction of sure winners) at time $\hat{t}_{k,N}$ in each state satisfies
        \begin{eqnarray*}
        \frac{H(\hat{t}_{k,N})}{N}\xrightarrow{p}\rho_{H}(1-e^{-\lambda \hat{t}_{k}}),\\
        \frac{L(\hat{t}_{k,N})}{N}\xrightarrow[]{p}\rho_{L}(1-e^{-\lambda \hat{t}_{k}}).
        \end{eqnarray*}
The sequence of equilibria aggregates information if
        \begin{equation*}
        \rho_{H}(1-e^{-\lambda \hat{t}_{k}})>k>\rho_{L}(1-e^{-\lambda \hat{t}_{k}}).
        \end{equation*}      
\cref{fig:2} depicts the quantities $\rho_{H}(1-e^{-\lambda \hat{t}_{k}})$ and $\rho_{L}(1-e^{-\lambda \hat{t}_{k}})$ as functions of $k$.

By \cref{prop:2}, the behavior of $\hat{t}_k$ depends on whether $k<\bar{k}$ or $k \geq \bar{k}$.
We consider each case separately.

\noindent \underline{Case 1: $k<\bar{k}$}. In this case, by \cref{prop:2}, $\hat{t}_k<\bar{t}$. We need to show that
\begin{equation*}
    \rho_{H}(1-e^{-\lambda \hat{t}_{k}})>k>\rho_{L}(1-e^{-\lambda \hat{t}_{k}}).
\end{equation*}
The proof is similar to that for information aggregation in a static voting model (see e.g.~\citet{feddersen1997voting}, \citet{duggan2001bayesian}). Recall that when the number of agents is large, each unsure voter at time $\hat{t}_k$ behaves like a single myopic decision-maker; thus, by $\hat{t}_k<\bar{t}$ and \eqref{eqn:1}, she would prefer $R$ if she knew the state was $H$ and $S$ if she knew the state was $L$. Now, if  
\begin{equation*}
    k\geq\rho_{H}(1-e^{-\lambda \hat{t}_{k}})>\rho_{L}(1-e^{-\lambda \hat{t}_{k}}),
\end{equation*}
then, conditional on being pivotal at time $\hat{t}_k$, an unsure voter would infer that the state must be $H$, and would thus prefer $R$. Similarly, if 
\begin{equation*}
    \rho_{H}(1-e^{-\lambda \hat{t}_{k}})>\rho_{L}(1-e^{-\lambda \hat{t}_{k}})\geq k,
\end{equation*}
then, conditional on being pivotal at time $\hat{t}_k$, an unsure voter would infer that the state must be $L$, and would thus prefer $S$. In either case, she would not be indifferent at time $\hat{t}_k$.

\noindent \underline{Case 2: $k\geq\bar{k}$}. In this case, by \cref{prop:2}, $\hat{t}_k = \bar{t}$, so the vote share of $R$ at time $\hat{t}_k$ is constant:
\begin{equation*}
        \rho_{H}(1-e^{-\lambda \hat{t}_{k}})=\rho_{H}(1-e^{-\lambda \overline{t}})=\frac{\rho_H g-s}{g-s}\;\forall k\geq\bar{k}.
\end{equation*}
In particular, when the state is $H$, information is not aggregated for any $k\in[\frac{\rho_H g-s}{g-s},1)$.

Note that when $k\geq\bar{k}$, in the limit as $N\to\infty$, each unsure voter becomes certain that the realized state is $H$ (conditional on being pivotal at $\bar{t}$):
\begin{equation*}
    \lim_{N\to\infty}\frac{Pr(H|kN-1,\bar{t})}{Pr(L|kN-1,\bar{t})}=\infty\;\,\forall k>\bar{k}.
\end{equation*}
However, she votes for $S$ at $\bar{t}$ since she is pessimistic about her type. 

The inefficiency in this case is induced by asymmetry in the agents' uncertainty about their types: At $\hat{t}_k$, sure winners are certain that their types are good, while unsure voters are uncertain about their types and hold pessimistic beliefs. This asymmetry creates a bias towards $S$: Those of the unsure voters whose types are good would vote for $R$ if they knew their types, but since they are uncertain, they vote for $S$ even if they are sure the state is $H$. Thus, $S$ is implemented in state $H$ whenever $k>\frac{\rho_H g-s}{g-s}$. 

These observations connect our work to that of \citet{fernandez1991resistance} and \citet{ali2025political}, who also show that asymmetric uncertainty may impair collective decision-making. Their papers consider a static voting model in which both the voting rule and the asymmetric uncertainty are exogenous. In the present paper, by contrast, the asymmetric uncertainty is endogenized by the collective decision rule: Raising the threshold $k$ leads to increased experimentation and thus induces more asymmetric uncertainty.

By \cref{prop:3}, we can approximate the first-best outcome arbitrarily closely by choosing a sufficiently small $k$, since $\lim_{k\to0}\hat{t}_{k}=0$. That is, when $k$ is small enough, information aggregation occurs almost immediately. However, information is not aggregated under the unanimity rule, under which $R$ is implemented if at least one agent votes for it, and $S$ is implemented if all the agents vote for it.\footnote{If $R$ is always implemented until some time $t$, then the probability that there is no sure winner in state $\omega\in\{H,L\}$ is $[1-\rho_{\omega}(1-e^{-\lambda t})]^N$. Information is aggregated if there exists $\hat{t}(N)$ such that $\lim_{N\to\infty}[1-\rho_{H}(1-e^{-\lambda \hat{t}(N)})]^N=0$ and $\lim_{N\to\infty}[1-\rho_{L}(1-e^{-\lambda \hat{t}(N)})]^N=1$. However, these two equations cannot hold at the same time.}


\section{Reversible Case}
\label{sec:4}

In our basic model, the safe action $S$ is irreversible; that is, once implemented, it remains fixed. We now analyze a variation of the model in which $S$ is reversible. As before, we examine perfect Bayesian equilibria in which (\romannumeral 1) sure winners always vote for $R$, and (\romannumeral 2) all unsure voters use the same pure strategy.

Let $H^t$ denote the public history at $t$, i.e., the number of votes cast for $R$ at each time from $0$ to $t$. Let $\mathcal{H}^t$ denote the set of all possible histories $H^t$, and let
\begin{equation*}
    \mathcal{H}=\bigcup_{t\in[0,\infty)}\mathcal{H}^t.
\end{equation*}
The strategy of the unsure voters is then characterized by a function
\begin{equation*}
    d:\mathcal{H}\to\{R,S\}.
\end{equation*}

In the equilibrium of the basic model with irreversible $S$, the number of votes for $R$ up to the equilibrium cut-off time carries no information; the unsure voters can learn from the other agents' payoffs only by conditioning on being pivotal. However, when $S$ is reversible, the unsure voters can all vote for $S$, observe the number of sure winners, and use this information to time their subsequent votes for $S$. Thus, reversibility enables the unsure voters to communicate through their voting patterns.

Given a history $H^t$, let $W(H^t)$ denote the set of possibilities for the number of sure winners at $t$. For example, fixing some $t'<t''$, suppose all unsure voters vote for $S$ at time $t'$. Also suppose there are $K$ sure winners at time $t'$, and the unsure voters all vote for $R$ at each $t\in(t',t'')$. Then
\begin{equation*}
    W(H^{t'})=K,
\end{equation*}
\begin{equation*}
    W(H^t)=\{K,K+1,\ldots,N\}\;\,\forall t\in(t',t'').
\end{equation*}

Now, consider an arbitrary equilibrium. Let $V(H^t)$ denote an unsure voter's discounted expected payoff from the equilibrium at time $t$ given the history $H^t$. Then
\begin{equation*}
    V(H^t)=\sum_{K\in W(H^t)}Pr(K|H^t)V(H^t,K),
\end{equation*}
where $Pr(K|H^t)$ is an unsure voter's belief that there are $K$ sure winners given $H^t$, and $V(H^t,K)$ is an unsure voter's discounted expected payoff from the equilibrium at $t$, given $H^t$ and given that there are $K$ sure winners. 

As before, we analyze equilibria in undominated strategies.
\begin{definition}
    \label{def:2}
    An equilibrium is an \emph{equilibrium in undominated strategies} if for each $t>0$ and each $H^t\in\mathcal{H}^t$, when
    \begin{equation*}
        V(H^t,K)\leq \frac{s}{r}\;\,\forall K\in W(H^t)\bigcap\{1,\ldots,kN-1\},
    \end{equation*}
    the unsure voters' strategy satisfies $d(H^t)= S$.
\end{definition}
When there are more than $kN$ sure winners, the risky action $R$ is always implemented and the unsure voters' choices have no influence. The refinement in Definition 2 rules out certain trivial equilibria---namely, those in which, at some time $t$, an unsure voter votes for $R$ because all of the other agents are voting for $R$ and so her vote is non-pivotal, even though she would prefer to implement $S$ for the remaining periods whenever there are fewer than $kN$ sure winners. This refinement is closely related to that of \citet{strulovici2010learning}, and to other refinements which eliminate conditionally weakly dominated strategies.

The equilibrium in the basic model (characterized by a cut-off $\hat{t}_{k,N}$) remains an equilibrium in undominated strategies when $S$ is reversible. Furthermore, we can show that in each equilibrium in undominated strategies, all unsure voters must vote for $S$ after time $\hat{t}_{k,N}$. This is because, when there are fewer than $kN$ sure winners, the unsure voters are the most optimistic about their type when there are exactly $kN-1$ sure winners. Therefore, by \cref{prop:3},
\begin{proposition}
    \label{prop:4}
    There is no sequence of equilibria in undominated strategies aggregating information when $k\geq\frac{\rho_Hg-s}{g-s}$.
\end{proposition}

\cref{prop:4} reinforces our argument that the collective decision rule should favor experimentation.

Finally, we can always construct a sequence of equilibria in undominated strategies that fails to aggregate information for all $k\in(0,1]$. Consider the following equilibrium: all unsure voters vote for $R$ at each time $t<t_1$, and they vote for $S$ at $t_1$. If there are no sure winners at time $t_1$, then all unsure voters vote for $S$ in all subsequent periods. If there are $K\geq 1$ sure winners at time $t_1$, then all unsure voters vote for $R$ until time $\hat{t}_{k,N}$ and for $S$ thereafter.\footnote{The unsure voters switch to $S$ at time $\hat{t}_{k,N}$ regardless of the number of sure winners at $t_1$, because they make their decisions conditional only on the event that there are $kN-1$ sure winners at $\hat{t}_{k,N}$; the times at which those sure winners received their lump sums are irrelevant. Equivalently, one could analyze the equilibrium in the setting where $S$ is irreversible and the agents observe when the first sure winner emerges.} We can show that along the sequence of these equilibria, $S$ is implemented in state $H$ with strictly positive probability.


\section{Related Literature}
\label{sec:5}

This paper contributes in three ways to the literature on experimentation with multiple agents, initiated by \citet{bolton1999strategic} and \citet{keller2005strategic}. First, we study a setting in which the agents jointly make a collective decision, rather than one in which each agent makes an individual decision. Second, we examine a form of inefficiency that arises from over-experimentation, emphasizing the effect of asymmetric uncertainty, whereas previous research mostly focuses on inefficiency due to under-experimentation induced by free-riding. Third, the existing literature typically assumes that the agents' payoffs are publicly observed, so that agents can learn directly from each other if their payoffs are correlated. In contrast, we assume each agent’s payoff is privately observed, so that agents learn from each other indirectly, by conditioning on being pivotal. Thus, different collective decision rules induce different incentives for experimentation. In a similar spirit, \citet{halac2017contests} analyze contests designed to encourage experimentation, in which each agent indirectly learns from the others by conditioning on the continuation of the contest. \citeauthor{halac2017contests} show that the principal can create stronger incentives for experimentation by committing to share the prize among all successful agents, rather than rewarding only the first success. 

Our paper is also related to the literature on information aggregation across strategic voters, initiated by \citet{austen1996information} and \citet{feddersen1997voting,feddersen1998convicting}. Most works in this field analyze static models in which the voters' preferences are exogenous. Voters receive private information about an unknown state that affects all of their payoffs in the same direction. By contrast, we analyze a dynamic model in which the agents receive increasingly precise information about both their preferences and the unknown state as experimentation proceeds. Notably, we show that the availability of this more precise information may reduce total welfare if it increases heterogeneity among the agents.

\citet{chan2018deliberating}, \citet{gieczewski2021policy}, and \citet{gieczewski2024experimentation} also examine collective experimentation. Their papers focus on settings in which agents hold different preferences ex ante, but information revealed through experimentation tends to reconcile them. By contrast, in our setting, the agents have ex ante common interests, and information revealed through experimentation causes their preferences to diverge.

In addition, our work is closely related to that of \citet{sun2023social}, in which agents have independent types and receive private signals during experimentation. \citeauthor{{sun2023social}} study the problem of a planner seeking a welfare-maximizing collective decision rule. \citet{moldovanu2021brexit} consider a setting in which the agents jointly decide between a reversible option and an irreversible one. They show that the collective decision rule should favor the reversible option; otherwise, the agents' failure to coordinate may diminish the option value of the reversible option. Finally, our paper shares several features with that of \citet{murto2011learning}, who analyze information aggregation in a stopping game with uncertain payoffs that are correlated among players. In their model, players make individual decisions about when to exit the game, whereas in our model the agents choose between two actions collectively.

\section{Concluding Remarks}
\label{sec:6}

This paper studies a dynamic model of collective decision-making in which the agents jointly decide whether to undertake a risky action. We show how strategic voting shapes the agents' incentives for experimentation and provide conditions under which information is aggregated. 

In our model, each agent's payoff stream is privately observed. For the case in which each agent's payoff stream is publicly observed, it is straightforward to extend the analysis in \citet{strulovici2010learning} to show that there is always a unique Markov equilibrium in undominated strategies. However, we cannot resolve whether information is aggregated along the sequence of those equilibria. The issue is that the equilibrium is characterized by numerous cut-offs when the number of agents grows large, and we do not have a closed-form expression for those cut-offs. We can show only that in state $L$, when the number of agents is large, the safe action $S$ is implemented almost immediately.

The results in this paper suggest several promising directions for future research. For example, we have assumed that both the state and the agents' types are fixed. It would be interesting to allow them to evolve with time, leading to richer dynamics and interactions among the agents. It may also be worthwhile to examine a model in which the speed of experimentation depends on the number of votes in favor of experimentation. The study of such a model would shed light on the collaboration problems analyzed by \citet{bonatti2011collaborating} and \citet{campbell2014delay}.

\begin{appendix}

\section{Proofs}

\subsection{Proof of \cref{prop:1}}
\label{proof1}

Rewrite \eqref{eqn:5} as:
\begin{equation}
    \label{eqn:6}
    s=Pr(good|kN-1,\hat{t}_{k,N})\lambda\left[z+\frac{g}{r}-\frac{s}{r}+(N-kN)(\frac{Pr(good|kN,\hat{t}_{k,N})g}{r}-\frac{s}{r})\right].
\end{equation}
Let $f(\hat{t}_{k,N})$ denote the right side of \eqref{eqn:6}. By \eqref{eqn:3},
\begin{equation*}
    \lim_{t\to 0}\frac{Pr(H|kN,t)}{Pr(L|kN,t)}>\frac{q_0}{1-q_0},
\end{equation*}
\begin{equation*}
    \lim_{t\to 0}\frac{Pr(H|kN-1,t)}{Pr(L|kN-1,t)}\geq\frac{q_0}{1-q_0}.
\end{equation*}
Hence, by \eqref{eqn:4},
\begin{equation*}
    \lim_{t\to 0}Pr(good|kN,t)>q_0\rho_H+(1-q_0)\rho_L,
\end{equation*}
\begin{equation*}
    \lim_{t\to 0}Pr(good|kN-1,t)\geq q_0\rho_H+(1-q_0)\rho_L.
\end{equation*}
Therefore, by \eqref{eqn:2},
\begin{equation*}
    \lim_{t\to 0}Pr(good|kN,t)g>s,
\end{equation*}
\begin{equation*}
    \lim_{t\to 0}Pr(good|kN-1,t)g\geq s.
\end{equation*}
This establishes:
\begin{equation}
\label{eqn:7}
    \lim_{t\to 0}f(t)>s.
\end{equation}
By \eqref{eqn:4},
\begin{equation*}
    \lim_{t\to \infty}Pr(good|kN,t)=\lim_{t\to \infty}Pr(good|kN-1,t)\leq\lim_{t\to \infty}Pr(good|H,t)=0.
\end{equation*}
Hence,
\begin{equation}
\label{eqn:8}
    \lim_{t\to \infty} f(t)=0.
\end{equation}

As discussed in \cref{sec:2.3}, $Pr(good|kN,t)$ and $Pr(good|kN-1,t)$ are both strictly decreasing in $t$. Hence, $f(t)$ is strictly decreasing in $t$ when $f(t)>0$. Since $f(t)$ is continuous in $t$ and satisfies \eqref{eqn:7} and \eqref{eqn:8}, the intermediate value theorem guarantees a unique $\hat{t}_{k,N}$ that solves \eqref{eqn:6} and characterizes the equilibrium.


\subsection{Proof of \cref{prop:2}}
\label{proof2}

We rewrite \eqref{eqn:5} by substituting \eqref{eqn:4}. The equilibrium cut-off $\hat{t}_{k,N}$ is the solution of
\begin{equation}
\label{eqn:9}
\frac{Pr(H|kN-1,\hat{t}_{k,N})}{Pr(L|kN-1,\hat{t}_{k,N})}\cdot\frac{v_H(k,N,\hat{t}_{k,N})}{v_L(k,N,\hat{t}_{k,N})}=1,
\end{equation}
where $v_H(k,N,\hat{t}_{k,N})$ is the pivotal unsure voter's gain in state $H$ by choosing $R$ instead of $S$,
\begin{equation*}
\begin{aligned}
    v_H(k,N,\hat{t}_{k,N})=&Pr(good|H,\hat{t}_{k,N})g-s+Pr(good|H,\hat{t}_{k,N})\lambda(\frac{g}{r}-\frac{s}{r})\\
    &+(N-kN)Pr(good|H,\hat{t}_{k,N})\lambda\left[\frac{Pr(good|H,\hat{t}_{k,N})g}{r}-\frac{s}{r}\right],
\end{aligned}   
\end{equation*}
and  $v_L(k,N,\hat{t}_{k,N})$ is the pivotal unsure voter's loss in state $L$ by choosing $R$ instead of $S$,
\begin{equation*}
\begin{aligned}
    v_L(k,N,\hat{t}_{k,N})=&s-Pr(good|L,\hat{t}_{k,N})g-Pr(good|L,\hat{t}_{k,N})\lambda(\frac{g}{r}-\frac{s}{r})\\
    &-(N-kN)Pr(good|L,\hat{t}_{k,N})\lambda\left[\frac{Pr(good|L,\hat{t}_{k,N})g}{r}-\frac{s}{r}\right].
\end{aligned}   
\end{equation*}

By \eqref{eqn:2},
\begin{equation}
    \label{eqn:10}
    \frac{Pr(H|kN-1,t)}{Pr(L|kN-1,t)}=\frac{q_0}{1-q_0}\cdot\frac{\rho_H}{\rho_L}\frac{1-\rho_L(1-e^{-\lambda t})}{1-\rho_H(1-e^{-\lambda t})}\cdot\left\{\left(\frac{\rho_H}{\rho_L}\right)^k\left[\frac{1-\rho_H(1-e^{-\lambda t})}{1-\rho_L(1-e^{-\lambda t})}\right]^{1-k}\right\}^{N}.
\end{equation}
For each $k\in(0,1)$, the fraction
\begin{equation*}
    \left(\frac{\rho_H}{\rho_L}\right)^k\left[\frac{1-\rho_H(1-e^{-\lambda t})}{1-\rho_L(1-e^{-\lambda t})}\right]^{1-k}
\end{equation*}
is strictly decreasing in $t$. Let
\begin{equation*}
    \hat{k}=\frac{\ln\frac{1-\rho_L}{1-\rho_H}}{\ln\frac{\rho_H}{\rho_L}+\ln\frac{1-\rho_L}{1-\rho_H}}.
\end{equation*}
When $k\geq\hat{k}$, we have:
\begin{equation*}
    \left(\frac{\rho_H}{\rho_L}\right)^k\left[\frac{1-\rho_H(1-e^{-\lambda t})}{1-\rho_L(1-e^{-\lambda t})}\right]^{1-k}>1 \;\,\forall t\geq 0.
\end{equation*}
When $k<\hat{k}$, the equation
\begin{equation*}
    \left(\frac{\rho_H}{\rho_L}\right)^k\left[\frac{1-\rho_H(1-e^{-\lambda t})}{1-\rho_L(1-e^{-\lambda t})}\right]^{1-k}=1
\end{equation*}
admits a unique positive solution in $t$. Let $\hat{t}'_k$ denote this solution. This solution is strictly increasing in $k$ with
\begin{equation*}
    \lim_{k\to 0}\hat{t}'_k=0,
\end{equation*}
\begin{equation*}
    \lim_{k\to \hat{k}}\hat{t}'_k=\infty.
\end{equation*}
There exists a unique $\bar{k}\in(0,\hat{k})$ such that $\hat{t}'_{\bar{k}}=\bar{t}$.

We argue that for each $k<\bar{k}$, the sequence of equilibrium cut-offs $\{\hat{t}_{k,N}\}_{N=1}^{\infty}$ converges to $\hat{t}'_k$. If not, we can find a subsequence $\{n_i\}_{i=1}^{\infty}$ of $\{1,2,\ldots\}$ such that either
\begin{equation}
\label{eqn:11}
    \lim_{i\to\infty}\hat{t}_{k,n_i}<\hat{t}'_k,
\end{equation}
or
\begin{equation}
\label{eqn:12}
    \lim_{i\to\infty}\hat{t}_{k,n_i}>\hat{t}'_k.
\end{equation}
In the first case with \eqref{eqn:11}, 
\begin{equation*}
    \lim_{i\to\infty}\left(\frac{\rho_H}{\rho_L}\right)^k\left[\frac{1-\rho_H(1-e^{-\lambda \hat{t}_{k,n_i}})}{1-\rho_L(1-e^{-\lambda \hat{t}_{k,n_i}})}\right]^{1-k}>1.
\end{equation*}
Hence,
\begin{equation*}
    \lim_{i\to\infty}\frac{Pr(H|kN-1,\hat{t}_{k,n_i})}{Pr(L|kN-1,\hat{t}_{k,n_i})}=\infty.
\end{equation*}
Furthermore, since
\begin{equation*}
    \lim_{i\to\infty}\hat{t}_{k,n_i}<\hat{t}'_k<\bar{t},
\end{equation*}
we can find $c>0$ such that
\begin{equation*}
    \lim_{i\to\infty}\frac{v_H(k,N,\hat{t}_{k,n_i})}{v_L(k,N,\hat{t}_{k,n_i})}=c.
\end{equation*}
Therefore,
\begin{equation*}
    \lim_{i\to\infty}\frac{Pr(H|kN-1,\hat{t}_{k,n_i})}{Pr(L|kN-1,\hat{t}_{k,n_i})}\cdot\frac{v_H(k,N,\hat{t}_{k,n_i})}{v_L(k,N,\hat{t}_{k,n_i})}=\infty,
\end{equation*}
which contradicts \eqref{eqn:9}. In the second case with \eqref{eqn:12}, we can show:
\begin{equation*}
    \lim_{i\to\infty}\frac{Pr(H|kN-1,\hat{t}_{k,n_i})}{Pr(L|kN-1,\hat{t}_{k,n_i})}=0,
\end{equation*}
and
\begin{equation*}
    \lim_{i\to\infty}\frac{v_H(k,N,\hat{t}_{k,n_i})}{v_L(k,N,\hat{t}_{k,n_i})}<\infty.
\end{equation*}
Therefore,
\begin{equation*}
    \lim_{i\to\infty}\frac{Pr(H|kN-1,\hat{t}_{k,n_i})}{Pr(L|kN-1,\hat{t}_{k,n_i})}\cdot\frac{v_H(k,N,\hat{t}_{k,n_i})}{v_L(k,N,\hat{t}_{k,n_i})}=0,
\end{equation*}
which contradicts \eqref{eqn:9}.

We argue that for each $k\geq\bar{k}$, the sequence of equilibrium cut-offs $\{\hat{t}_{k,N}\}_{N=1}^{\infty}$ converges to $\bar{t}$. If not, we can find a subsequence $\{n_i\}_{i=1}^{\infty}$ of $\{1,2,\ldots\}$ such that either
\begin{equation}
\label{eqn:13}
    \lim_{i\to\infty}\hat{t}_{k,n_i}<\bar{t},
\end{equation}
or
\begin{equation}
\label{eqn:14}
    \lim_{i\to\infty}\hat{t}_{k,n_i}>\bar{t}.
\end{equation}
The first case with \eqref{eqn:13} is similar to the case with \eqref{eqn:11}. We can show:
\begin{equation*}
    \lim_{i\to\infty}\frac{Pr(H|kN-1,\hat{t}_{k,n_i})}{Pr(L|kN-1,\hat{t}_{k,n_i})}\cdot\frac{v_H(k,N,\hat{t}_{k,n_i})}{v_L(k,N,\hat{t}_{k,n_i})}=\infty,
\end{equation*}
which leads to a contradiction. For the second case with \eqref{eqn:14}, note that for each $\hat{t}>\bar{t}$,
\begin{equation*}
    Pr(good|L,\hat{t})g<Pr(good|H,\hat{t})g<s.
\end{equation*}
Hence,
\begin{equation*}
    \lim_{i\to\infty}\frac{v_H(k,N,\hat{t}_{k,n_i})}{v_L(k,N,\hat{t}_{k,n_i})}<0,
\end{equation*}
and
\begin{equation*}
    \lim_{i\to\infty}\frac{Pr(H|kN-1,\hat{t}_{k,n_i})}{Pr(L|kN-1,\hat{t}_{k,n_i})}\cdot\frac{v_H(k,N,\hat{t}_{k,n_i})}{v_L(k,N,\hat{t}_{k,n_i})}\leq 0,
\end{equation*}
which contradicts \eqref{eqn:9}.

Therefore, 
\begin{equation*}
    \hat{t}_k=
    \begin{cases}
        \hat{t}'_k,\,\,\text{if}\,\,k<\bar{k},\\
        \bar{t},\,\,\,\,\,\text{if}\,\,k\geq\bar{k}.
    \end{cases}
\end{equation*}


\subsection{Proof of \cref{prop:3}}
\label{proof3}

By the law of large numbers and \cref{prop:2}, the vote share of $R$ (i.e., the fraction of sure winners) at time $\hat{t}_{k,N}$ in each state satisfies
        \begin{equation*}
       \frac{H(\hat{t}_{k,N})}{N}\xrightarrow{p}\rho_{H}(1-e^{-\lambda \hat{t}_{k}}),
       \end{equation*}
        \begin{equation*}
        \frac{L(\hat{t}_{k,N})}{N}\xrightarrow[]{p}\rho_{L}(1-e^{-\lambda \hat{t}_{k}}).
        \end{equation*}
The sequence of equilibria aggregates information if
        \begin{equation}
        \label{eqn:15}
        \rho_{H}(1-e^{-\lambda \hat{t}_{k}})>k>\rho_{L}(1-e^{-\lambda \hat{t}_{k}}).
        \end{equation}      

We first prove that \eqref{eqn:15} holds for each $k\leq\bar{k}$. Note that $\bar{k}$ is characterized in the \hyperref[proof2]{Proof of Proposition 2}. If \eqref{eqn:15} does not hold, consider the case in which
\begin{equation*}
     \rho_{H}(1-e^{-\lambda \hat{t}_{k}})>\rho_{L}(1-e^{-\lambda \hat{t}_{k}})\geq k.
\end{equation*}
The function $x^k(1-x)^{1-k}$ is strictly increasing in $x$ when $x\in[0,k)$ and strictly decreasing in $x$ when $x\in [k,1]$. Hence,
\begin{equation*}
    \left[\rho_{H}(1-e^{-\lambda \hat{t}_{k}})\right]^k\left[1-\rho_{H}(1-e^{-\lambda \hat{t}_{k}})\right]^{1-k}<\left[\rho_{L}(1-e^{-\lambda \hat{t}_{k}})\right]^k\left[1-\rho_{L}(1-e^{-\lambda \hat{t}_{k}})\right]^{1-k}.
\end{equation*}
Thus,
\begin{equation*}
    \left(\frac{\rho_H}{\rho_L}\right)^k\left[\frac{1-\rho_H(1-e^{-\lambda \hat{t}_k})}{1-\rho_L(1-e^{-\lambda \hat{t}_k})}\right]^{1-k}>1.
\end{equation*}
However, by the \hyperref[proof2]{Proof of Proposition 2}, the limit cut-off $\hat{t}_k$ must satisfy
\begin{equation*}
    \left(\frac{\rho_H}{\rho_L}\right)^k\left[\frac{1-\rho_H(1-e^{-\lambda \hat{t}_k})}{1-\rho_L(1-e^{-\lambda \hat{t}_k})}\right]^{1-k}=1,
\end{equation*}
which leads to a contradiction. The case in which
\begin{equation*}
     k\geq \rho_{H}(1-e^{-\lambda \hat{t}_{k}})>\rho_{L}(1-e^{-\lambda \hat{t}_{k}})
\end{equation*}
is similar.

For each $k\geq\bar{k}$, it follows that $\hat{t}_k=\bar{t}$ by \cref{prop:2}. Since information is aggregated when $k=\bar{k}$, we have:
\begin{equation*}
    \rho_{H}(1-e^{-\lambda \bar{t}})>\bar{k}>\rho_{L}(1-e^{-\lambda \bar{t}}).
\end{equation*}
Hence, information is aggregated in state $L$ for each $k>\bar{k}$:
\begin{equation*}
    \rho_{L}(1-e^{-\lambda \hat{t}_{k}})=\rho_{L}(1-e^{-\lambda \overline{t}})<\bar{k}<k.
\end{equation*}
However, in state $H$,
\begin{equation*}
    \rho_{H}(1-e^{-\lambda \hat{t}_{k}})=\rho_{H}(1-e^{-\lambda \overline{t}})=\frac{\rho_H g-s}{g-s}.
\end{equation*}
Therefore, information is aggregated when $k<\frac{\rho_H g-s}{g-s}$ and not aggregated when $k\geq\frac{\rho_H g-s}{g-s}$.

\subsection{Proof of \cref{prop:4}}
\label{proof4}

We now show that in each equilibrium in undominated strategies, unsure voters must vote for $S$ after $\hat{t}_{k,N}$. First, let $\tilde{t}$ be the unique solution to
\begin{equation*}
    Pr(good|H,\tilde{t})=\frac{rs}{(r+\lambda)g-\lambda s}.
\end{equation*}
Consider a version of our game involving only a single agent, who chooses between $R$ and $S$ at each instant. Assume that this agent is certain that the state is $H$ but uncertain about her type. Her optimal strategy is to choose $R$ until $\tilde{t}$ and then choose $S$ if no lump sum has arrived. An unsure voter’s expected payoff at each time in every equilibrium is lower than that of the single agent described above. Hence, in each equilibrium, for every $t\geq \tilde{t}$ and $H^t\in\mathcal{H}^t$,
\begin{equation*}
    V(H^t,K)\leq \frac{s}{r}\;\,\forall K\in W(H^t)\bigcap\{1,...,kN-1\}.
\end{equation*}
Therefore, unsure voters must switch to $S$ whenever $t\geq\tilde{t}$.

Since unsure voters are indifferent between $R$ and $S$ at $\hat{t}_{k,N}$ conditional on the event that there are $kN-1$ sure winners, we have, for each $t\in[\hat{t}_{k,N},\tilde{t}]$ and $H^t\in\mathcal{H}^t$ with $kN-1\in W(H^t)$,
\begin{equation*}
    V(H^t,kN-1)\leq \frac{s}{r}.
\end{equation*}
When the number of sure winners $K=kN-2$, there exists $\delta>0$ such that for each $t\in[\tilde{t}-\delta,\tilde{t}]$ and $H^t\in\mathcal{H}^t$ with $kN-2\in W(H^t)$,
\begin{equation*}
    V(H^t,kN-2)\leq \frac{s}{r}.
\end{equation*}
It suffices to find $\delta>0$ such that, conditional on the event that there are $kN-2$ sure winners, an unsure voter obtains a higher expected payoff by switching to $S$ at $\tilde{t}-\delta$ than by continuing with $R$ until $\tilde{t}$. Note that choosing $S$ instead of $R$ generates higher flow payoffs when $t>\hat{t}_{k,N}$ since:
\begin{equation*}
    Pr(good|kN-2,t)<Pr(good|kN-1,\hat{t}_{k,N})<\frac{s}{g}.
\end{equation*}
The difference in the discounted flow payoffs from $\tilde{t}-\delta$ to $\tilde{t}$ is at least
\begin{equation*}
    D_1=\int_{0}^{\delta}e^{-rt}\left[s-Pr(good|kN-1,\hat{t}_{k,N})g\right]dt.
\end{equation*}
However, suppose voters continue with $R$ from $\tilde{t}-\delta$ to $\tilde{t}$. If at least two unsure voters receive lump sums in that interval, then $R$ is implemented for all subsequent periods, which generates a difference in the expected payoff of each unsure voter at $\tilde{t}$ compared to the case in which unsure voters switch to $S$ at $\tilde{t}-\delta$. This difference is at most
\begin{equation*}
    D_2=e^{-r\delta}\cdot\left(\frac{g}{r}-\frac{s}{r}\right)\cdot\left(1-e^{Pr(good|kN-1,\hat{t}_{k,N})\lambda\delta}\right)^2.
\end{equation*}
The second term $\frac{g}{r}-\frac{s}{r}$ bounds the jump in each unsure voter's discounted payoff, and the third term $\left(1-e^{Pr(good|kN-1,\hat{t}_{k,N})\lambda\delta}\right)^2$ bounds the probability that at least two unsure voters receive the lump sums. Since
\begin{equation*}
    \lim_{\delta\to 0}\frac{D_2}{D_1}=0,
\end{equation*}
we can find a $\delta>0$ such that, conditional on the event that there are $kN-2$ sure winners, unsure voters receive higher expected payoffs by switching to $S$ at $\tilde{t}-\delta$. Moreover, for each $t\in[\tilde{t}-\delta,\tilde{t}]$ and $H^t\in\mathcal{H}^t$, 
\begin{equation*}
    V(H^t,K)\leq \frac{s}{r}\;\,\forall K\in W(H^t)\bigcap\{1,...,kN-2\}.
\end{equation*}
Hence, unsure voters must switch to $S$ when $t\geq\tilde{t}-\delta$. Furthermore, since the choice of $\delta$ is independent of $\tilde{t}$ (the end point of the interval), we can repeat the argument above backward to conclude that for each $t\in[\hat{t}_{k,N},\tilde{t}]$ and $H^t\in\mathcal{H}^t$, 
\begin{equation*}
    V(H^t,K)\leq \frac{s}{r}\;\,\forall K\in W(H^t)\bigcap\{1,...,kN-1\}.
\end{equation*}

\end{appendix}



\begin{thebibliography}{}
%
\bibitem[\protect\citeauthoryear{Ali, Mihm, and Siga}{2025}]{ali2025political}
Ali, S. Nageeb, Maximilian Mihm, and Lucas Siga (2025),
``The Political Economy of Zero-Sum Thinking.''
\textit{Econometrica}, 93(1), 41--70.
\endbibitem

\bibitem[\protect\citeauthoryear{Austen-Smith and Banks}{1996}]{austen1996information}
Austen-Smith, David, and Jeffrey Banks (1996),
``Information aggregation, rationality, and the Condorcet jury theorem.''
\textit{American Political Science Review}, 90(1), 34--45.
\endbibitem


\bibitem[\protect\citeauthoryear{Bolton and Harris}{1999}]{bolton1999strategic}
Bolton, Patrick, and Christopher Harris (1999),
``Strategic experimentation.''
\textit{Econometrica}, 67(2), 349--374.
\endbibitem

\bibitem[\protect\citeauthoryear{Bonatti and Hörner}{2011}]{bonatti2011collaborating}
Bonatti, Alessandro, and Johannes Hörner (2011),
``Collaborating.''
\textit{American Economic Review}, 101(2), 632--663.
\endbibitem

\bibitem[\protect\citeauthoryear{Campbell, Ederer, and Spinnewijn}{2014}]{campbell2014delay}
Campbell, Arthur, Florian Ederer, and Johannes Spinnewijn (2014),
``Delay and deadlines: Freeriding and information revelation in partnerships.''
\textit{American Economic Journal: Microeconomics}, 6(2), 163--204.
\endbibitem

\bibitem[\protect\citeauthoryear{Chan et al.}{2018}]{chan2018deliberating}
Chan, Jimmy, Alessandro Lizzeri, Wing Suen, and Leeat Yariv (2018),
``Deliberating collective decisions.''
\textit{Review of Economic Studies}, 85(2), 929--963.
\endbibitem


\bibitem[\protect\citeauthoryear{Duggan and Martinelli}{2001}]{duggan2001bayesian}
Duggan, John, and César Martinelli (2001),
``A Bayesian model of voting in juries.''
\textit{Games and Economic Behavior}, 37(2), 259--294.
\endbibitem



\bibitem[\protect\citeauthoryear{Feddersen and Pesendorfer}{1997}]{feddersen1997voting}
Feddersen, Timothy, and Wolfgang Pesendorfer (1997),
``Voting behavior and information aggregation in elections with private information.''
\textit{Econometrica}, 65(2), 1029--1058.
\endbibitem

\bibitem[\protect\citeauthoryear{Feddersen and Pesendorfer}{1998}]{feddersen1998convicting}
Feddersen, Timothy, and Wolfgang Pesendorfer (1998),
``Convicting the innocent: The inferiority of unanimous jury verdicts under strategic voting.''
\textit{American Political Science Review}, 92(1), 23--35.
\endbibitem



\bibitem[\protect\citeauthoryear{Fernandez and Rodrik}{1991}]{fernandez1991resistance}
Fernandez, Raquel, and Dani Rodrik (1991),
``Resistance to reform: Status quo bias in the presence of individual-specific uncertainty.''
\textit{American Economic Review}, 81(5), 1146--1155.
\endbibitem




\bibitem[\protect\citeauthoryear{Gieczewski}{2021}]{gieczewski2021policy}
Gieczewski, Germán (2021),
``Policy persistence and drift in organizations.''
\textit{Econometrica}, 89(1), 251--279.
\endbibitem


\bibitem[\protect\citeauthoryear{Gieczewski and Kosterina}{2024}]{gieczewski2024experimentation}
Gieczewski, Germán, and Svetlana Kosterina (2024),
``Experimentation in Endogenous Organizations.''
\textit{Review of Economic Studies}, 91(3), 1711--1745.
\endbibitem

\bibitem[\protect\citeauthoryear{Halac, Kartik, and Liu}{2017}]{halac2017contests}
Halac, Marina, Navin Kartik, and Qingmin Liu (2017),
``Contests for experimentation.''
\textit{Journal of Political Economy}, 125(5), 1523--1569.
\endbibitem


\bibitem[\protect\citeauthoryear{Keller, Rady, and Cripps}{2005}]{keller2005strategic}
Keller, Godfrey, Sven Rady, and Martin Cripps (2005),
``Strategic experimentation with exponential bandits.''
\textit{Econometrica}, 73(1), 39--68.
\endbibitem






\bibitem[\protect\citeauthoryear{Moldovanu and Rosar}{2021}]{moldovanu2021brexit}
Moldovanu, Benny, and Frank Rosar (2021),
``Brexit: A comparison of dynamic voting games with irreversible options.''
\textit{Games and Economic Behavior}, 130(5), 85--108.
\endbibitem






\bibitem[\protect\citeauthoryear{Murto and Välimäki}{2011}]{murto2011learning}
Murto, Pauli, and Juuso Välimäki (2011),
``Learning and information aggregation in an exit game.''
\textit{Review of Economic Studies}, 78(4), 1426--1461.
\endbibitem




\bibitem[\protect\citeauthoryear{Strulovici}{2010}]{strulovici2010learning}
Strulovici, Bruno (2010),
``Learning while voting: Determinants of collective experimentation.''
\textit{Econometrica}, 78(3), 933--971.
\endbibitem






\bibitem[\protect\citeauthoryear{Sun, Thomas, and Yamashita}{2023}]{sun2023social}
Sun, Yiman, Caroline Thomas, and Takuro Yamashita (2023),
``Social Choice under Gradual Learning.''
Working Paper.
\endbibitem







\end{thebibliography}


\end{document}